\documentclass[aps,pra,twocolumn,showpacs,notitlepage,superscriptaddress,letterpaper]{revtex4-1}

\usepackage{graphicx, amsmath, amssymb, amsfonts, pifont, bm, cancel}
\usepackage[usenames]{color}
\usepackage{subfigure}
\usepackage[normalem]{ulem} 
\usepackage{circuitikz}
\usepackage{hyperref}
\usepackage{threeparttable}
\usepackage{array}
\newcolumntype{P}[1]{>{\centering\arraybackslash}p{#1}}
\newcolumntype{M}[1]{>{\centering\arraybackslash}m{#1}}

\definecolor{MyDarkBlue}{rgb}{0,0,1}

\newcommand{\beq}{\begin{equation}}
\newcommand{\eeq}{\end{equation}}
\newcommand{\bel}{\begin{align*}}
\newcommand{\tamam}{\end{align*}}

\newcommand{\ket}[1]{|#1\rangle}

\newcommand{\beqa}{\begin{eqnarray}}             
\newcommand{\eeqa}{\end{eqnarray}}               
\newcommand{\bra}[1]{\langle#1\vert}                 

\newcommand{\ahat}{\hat{a}_{\text{o}}}
\newcommand{\adag}{\hat{a}^{\dagger}_{\text{o}}}
\newcommand{\bhat}{\hat{b}_{\text{m}}}
\newcommand{\bdag}{\hat{b}^{\dagger}_{\text{m}}}

\newcommand{\sigmaeg}{\hat{\sigma}_{\text{eg}}}
\newcommand{\sigmage}{\hat{\sigma}_{\text{ge}}}

\newcommand{\omegam}{\omega_{\text{m}}}
\newcommand{\omegac}{\omega_{\text{c}}}
\newcommand{\ncav}{n_{\text{c}}}
\newcommand{\omegad}{\omega_{\text{drive}}}
\newcommand{\piezoacoustic}{piezo-acoustic}
\newcommand{\gom}{g_{\text{om}}}
\newcommand{\Gom}{G_{\text{om}}}
\newcommand{\gpe}{g_{\text{pe}}}
\newcommand{\kappaRO}{\kappa_{\text{RO}}}
\newcommand{\kappaeQ}{\kappa_{\text{e,q}}}
\newcommand{\kappaPQ}{\kappa_{\text{P,q}}}
\newcommand{\gRO}{g_{\text{RO}}}
\newcommand{\fRO}{f_{\text{RO}}}
\newcommand{\AJ}{A_{\text{J}}}
\newcommand{\kappaoe}{\kappa_{\text{e,o}}}
\newcommand{\kappaoi}{\kappa_{\text{i,o}}}
\newcommand{\kappao}{\kappa_{\text{o}}}
\newcommand{\kappami}{\kappa_{\text{i,m}}}
\newcommand{\kappam}{\kappa_{\text{m}}}
\newcommand{\kappamiedecay}{\kappa_{\text{m},T_{1}}}
\newcommand{\gammaom}{\gamma_{\text{om}}}
\newcommand{\Tf}{T_{\text{f}}}
\newcommand{\etae}{\eta_{\text{e,ro}}}
\newcommand{\etai}{\eta_{\text{i,ro}}}
\newcommand{\etat}{\eta_{\text{t}}}
\newcommand{\etaswap}{\eta_{\text{swap}}}
\newcommand{\etaRO}{\eta_{\text{ro}}}
\newcommand{\etasys}{\eta_{\text{sys}}}
\newcommand{\etacoupler}{\eta_{\text{cplr}}}
\newcommand{\etatran}{\eta_{\text{tran}}}
\newcommand{\etaSPD}{\eta_{\text{spd}}}
\newcommand{\nadd}{n_{\text{add}}}
\newcommand{\nm}{n_{\text{m}}}
\newcommand{\tauRO}{\tau_{\text{ro}}}
\newcommand{\Ppi}{P_{\pi}}
\newcommand{\Pnopi}{P_{0}}

\begin{document}

\title{Quantum transduction of optical photons from a superconducting qubit}
\author{Mohammad Mirhosseini}
\thanks{These authors contributed equally to this work.}
\author{Alp Sipahigil}
\thanks{These authors contributed equally to this work.}
\affiliation{Kavli Nanoscience Institute and Thomas J. Watson, Sr., Laboratory of Applied Physics, California Institute of Technology, Pasadena, California 91125, USA.}
\affiliation{Institute for Quantum Information and Matter, California Institute of Technology, Pasadena, California 91125, USA.}
\author{Mahmoud Kalaee}
\thanks{These authors contributed equally to this work.}
\affiliation{Kavli Nanoscience Institute and Thomas J. Watson, Sr., Laboratory of Applied Physics, California Institute of Technology, Pasadena, California 91125, USA.}
\affiliation{Institute for Quantum Information and Matter, California Institute of Technology, Pasadena, California 91125, USA.}
\affiliation{AWS Center for Quantum Computing, Pasadena, California 91125, USA.}
\author{Oskar Painter}
\email{opainter@caltech.edu}
\homepage{http://copilot.caltech.edu}
\affiliation{Kavli Nanoscience Institute and Thomas J. Watson, Sr., Laboratory of Applied Physics, California Institute of Technology, Pasadena, California 91125, USA.}
\affiliation{Institute for Quantum Information and Matter, California Institute of Technology, Pasadena, California 91125, USA.}
\affiliation{AWS Center for Quantum Computing, Pasadena, California 91125, USA.}

\date{\today}
\maketitle

\noindent\textbf{\noindent Bidirectional conversion of electrical and optical signals lies at the foundation of the global internet.  Such converters are employed at repeater stations to extend the reach of long-haul fiber optic communication systems and within data centers to exchange high-speed optical signals between computers. Likewise, coherent microwave-to-optical conversion of single photons would enable the exchange of quantum states between remotely connected superconducting quantum processors, a promising quantum computing hardware platform~\cite{Google2019}. Despite the prospects of quantum networking~\cite{Briegel1998}, maintaining the fragile quantum state in such a conversion process with superconducting qubits has remained elusive. Here we demonstrate the conversion of a microwave-frequency excitation of a superconducting transmon qubit into an optical photon. We achieve this using an intermediary nanomechanical resonator which converts the electrical excitation of the qubit into a single phonon by means of a piezoelectric interaction~\cite{OConnell2010}, and subsequently converts the phonon to an optical photon via radiation pressure~\cite{Meenehan2015}. We demonstrate optical photon generation from the qubit with a signal-to-noise greater than unity by recording quantum Rabi oscillations of the qubit through single-photon detection of the emitted light over an optical fiber. With proposed improvements in the device and external measurement set-up, such quantum transducers may lead to practical devices capable of realizing new hybrid quantum networks~\cite{Briegel1998,Muralidharan2016}, and ultimately, distributed quantum computers~\cite{Monroe2014,Fitzsimons2017}.}

Recent developments with superconducting qubits have demonstrated fast, high fidelity single- and two-qubit gates, making them one of the leading platforms for realizing quantum computers~\cite{Google2019}. While the low-loss environment of a superconductor and the strong single-photon nonlinearity from the Josephson effect provide an ideal combination for processing quantum information in the microwave domain~\cite{Devoret2013}, optical photons are a natural choice for quantum networking tasks~\cite{kimble2008quantum} where they provide low propagation loss in room-temperature environments~\cite{OBrien:2009eu}. A coherent microwave-to-optical interface can thus lead to hybrid architectures for quantum repeaters~\cite{Briegel1998,Muralidharan2016} by connecting superconducting qubits and ultra high-Q microwave cavities~\cite{Reagor:2013kf}-- serving as logic and memory registers-- to flying optical qubits as a means of long-distance information transfer. Although the process of frequency conversion can be understood simply as a noise-free and loss-less linear operation, an optical interface for superconducting qubits has not been realized to this date because of the technical challenges inherently set by the vast frequency difference of microwave ($\sim$5 GHz) and telecom-band optical ($\sim$200 THz) photons. 

Microwave-to-optical frequency conversion can be achieved by bulk optical nonlinearities~\cite{Fan2018}. Alternatively, effective nonlinearities can be realized by intermediary degrees of freedom such as rare earth ions, magnons, or phonons~\cite{Histaomi2016,OBrien2014,Lambert2019} that exhibit simultaneous electrical and optical susceptibilities. Employing engineered nanomechanical resonators as such intermediary channels has been a particularly promising direction, where the pioneering work with integrated electromechanical and optomechanical systems in the past decade has demonstrated electrical and optical preparation, control, and readout of mechanical modes near their quantum ground state~\cite{OConnell2010,Teufel2011,Chan2011}. These demonstrations, together with rapid developments in superconducting quantum circuits~\cite{Devoret2013}, have motivated recent experimental efforts to combine electromechanical and optomechanical devices to build a microwave-to-optical quantum transducer~\cite{Bochmann2013,Andrews2014,Bagci:2014du,Balram:2016eu,Forsch2018}. Although this approach has led to impressive conversion efficiencies~\cite{Higginbotham}, all demonstrations so far have been limited to classical signals due to a combination of challenges associated with optically-induced or thermal noise, small transduction bandwidths, and device integration complexities. 

\begin{figure*}
\begin{center}
\includegraphics[width=1\textwidth]{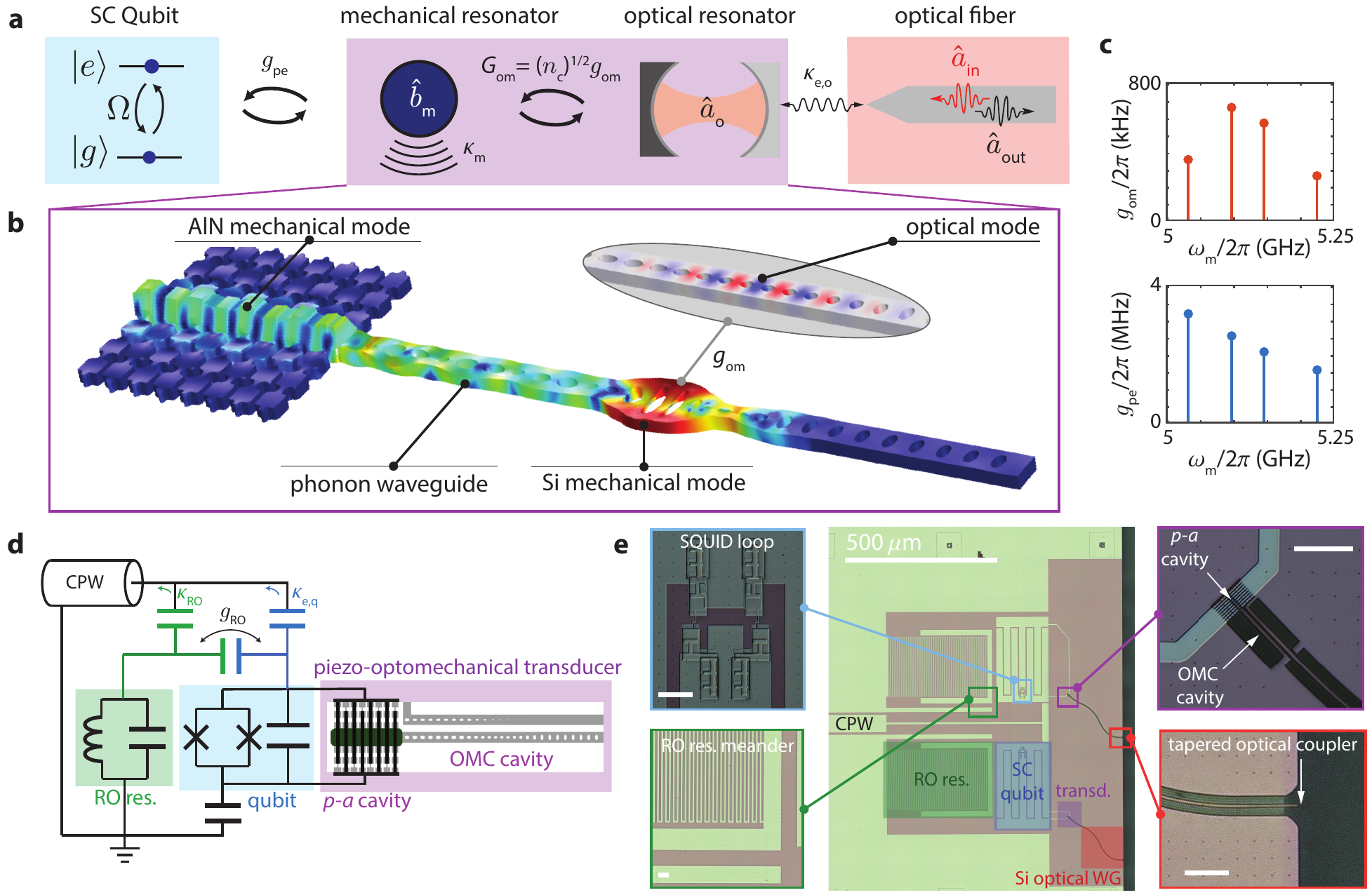}
\caption{\textbf{Quantum transducer setup.} \textbf{a}, Schematic of the microwave-to-optical transduction process. The mechanical mode ($\hat{b}_m$) couples to both the superconducting qubit ($\sigmage = \ket{g}\bra{e}$) and an optical mode ($\hat{a}_o$) via piezoelectric ($\gpe$) and optomechanical ($\gom$) vacuum coupling rates, respectively.  \textbf{b}, Numerically simulated resonant modes of the piezo-optomechanical transducer. \textbf{c}, Simulated vacuum optomechanical (top) and piezoelectric (bottom) coupling rates to the hybridized mechanical modes. \textbf{d}, Electrical circuit representation of the integrated qubit and transducer device.  \textbf{e}, Optical micrograph of a pair of fabricated devices, showing the readout resonator (green), transmon qubit (blue), transducer element (purple), and silicon out-coupling waveguide (red). Corresponding zoomed in optical images of the different device sections are shown to the left and right, with white scale bar corresponding to $10$ microns.  Electrical and optical readout is performed in reflection via the CPW and a lensed optical fiber (not shown; optically coupled to the tapered silicon waveguide coupler), respectively.  Figure labels: Josephson-junction superconducting quantum interference device (SQUID); read-out resonator (RO res.); superconducting transmon qubit (SC qubit); optomechanical crystal cavity (OMC cavity); \piezoacoustic{} ($p$-$a$) cavity; silicon optical waveguide (Si optical WG).}\label{fig:schematic} 
\end{center}
\end{figure*}

Here, we demonstrate optical photon generation from a superconducting qubit and detection over an optical fiber link. This is achieved using a monolithic platform that integrates a transmon qubit with piezo-optomechanical transducer components on the same microchip. In contrast to earlier work based on continuous wave linear transducers, we use a pulsed scheme where we coherently transfer the quantum state of a qubit into a nanomechanical mode via a piezoelectric swap operation, and subsequently convert it to the optical domain using a pulsed laser drive. This pulsed approach crucially separates the electric and optical parts of the sequence, making the experiment robust against light-induced noise in the superconducting circuitry, and eliminates the need for matching the interaction bandwidths (i.e. impedance matching) on the microwave and optical sides~\cite{Safavi-Naeini2011b}. 

Our device is based on a hybrid materials platform consisting of thin-film aluminum nitride (AlN) sputter deposited on a silicon-on-insulator wafer with a high resistivity silicon (Si) device layer; this materials system simultaneously provides for large piezoelectric and optomechanical coupling coefficients and low electromagnetic absorption in the microwave and (telecom) optical frequency bands~\cite{Meenehan2015}. Using this integrated device platform, we demonstrate the detection of single optical photons from a transmon qubit, and use them to register quantum Rabi oscillations of the qubit via an entirely optical measurement.  We use the transmon as a single photon source to directly quantify the overall added noise photons in the process, finding a value ($0.57\pm0.2$) below unity, which surpasses the threshold for faithful exchange of quantum information~\cite{Zeuthen2017}. We discuss practical avenues for achieving reduced noise and orders of magnitude improvement in the total system efficiency ($10^{-5}$), and outline a clear path to future experiments involving the remote entanglement of superconducting qubits via optical photons.

Figure~\ref{fig:schematic}a shows the schematic of the system in our experiment, where an intermediary mechanical mode is coupled to a qubit via a resonant piezoelectric interaction and to an optical mode via a parametric optomechanical interaction. The Hamiltonian for this system can be written as $\hat{H} = \hat{H}_0 + \hat{H}_\text{pe} + \hat{H}_\text{om}$. Here, $\hat{H}_0/\hbar = \omegac \adag\ahat + \omegam \bdag\bhat + \omega_\text{q}(t) \hat{\sigma}_{ee}$ describes the evolution of non-interacting sub-systems, with $\hat{H}_\text{pe}/\hbar =\gpe( \sigmaeg\bhat +\sigmage\bdag)$ and $\hat{H}_\text{om}/\hbar =\Gom(t)(\adag\bhat +\ahat\bdag)$ describing the piezoelectric and optomechanical interactions, respectively. Here, we use the `beam-splitter' form of the optomechanical interaction which is specific to the case where the optical drive is red-detuned from the optical cavity resonance ($\Delta = \omegac-\omegad = \omegam$) and assumes operation in the resolved-sideband limit, where $\omegam\gg \kappa_\text{o}$ ($\kappa_\text{o}$ is the linewidth of the optical resonance)~\cite{Aspelmeyer:2014ce}. In this case, the optomechanical coupling $\Gom(t) = \sqrt{n_\text{c}(t)} \gom$ is parametrically enhanced from the single-photon rate, $\gom$, by the intra-cavity photon number $n_\text{c}(t)$ from the pump laser at $\omegad$. 

Physical realization of the above Hamiltonian requires a material platform with simultaneous access to optomechanical and piezoelectric components. High fidelity qubit-mechanics swap operations can be realized based on the piezoelectric response of materials such as AlN and lithium niobate, where the electric field from a qubit can be transformed to displacement in a mechanical resonator~\cite{Chu:2017kt,Arrangoiz-Arriola}. Quantum coherent optomechanical readout of mechanical modes can also be realized in optomechanical crystal (OMC) cavities~\cite{Meenehan2015,Hong:2017id}, where co-localization of mechanical and optical fields in a patterned nanobeam structure results in a large parametric coupling caused by radiation pressure and photo-elastic effect. In our experiment, we use high resistivity silicon as the material of choice, which allows for integrating OMC cavities with large optomechanical coupling rates and low mechanical loss together with transmon qubits on the same substrate~\cite{Keller2017}. Silicon is not piezoelectric by symmetry of its lattice; however, by depositing and selectively patterning a thin-film of AlN on the Si substrate a localized piezoelectric response may be achieved (see Methods). 

\begin{figure}
\begin{center}

\includegraphics[width=1\columnwidth]{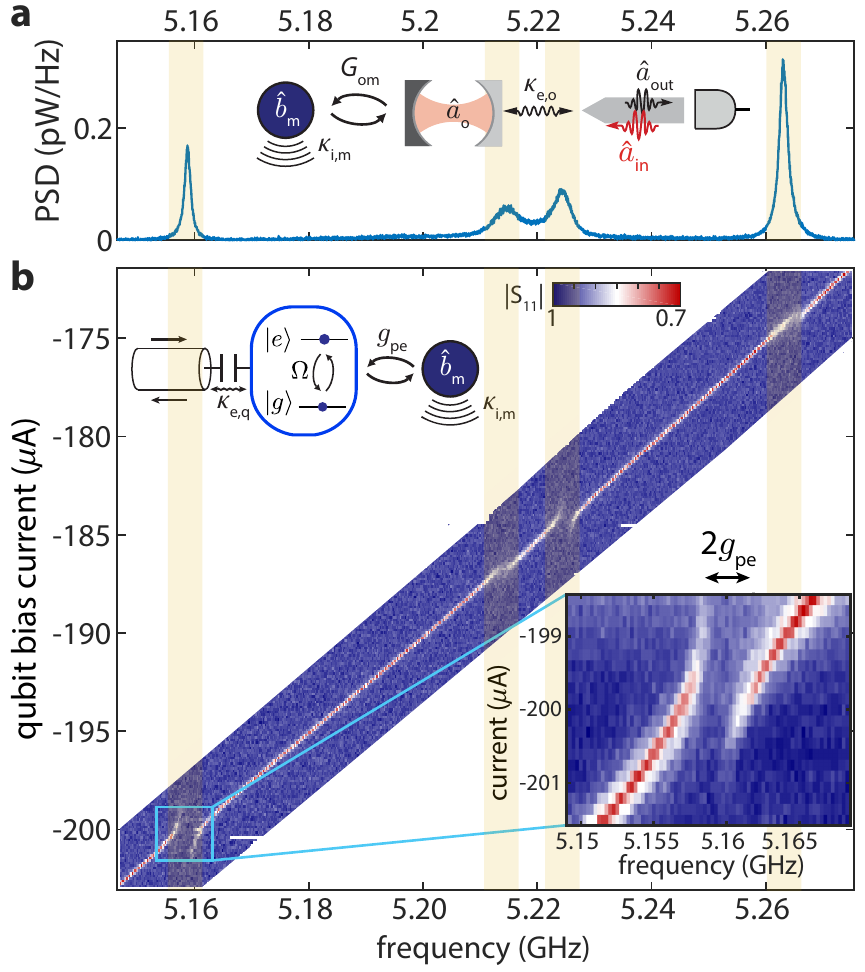}
\caption{\textbf{Microwave and Optical spectroscopy of mechanics.}  \textbf{a}, Optically transduced thermal mechanical spectrum as measured with the pump laser detuned by $\Delta/2\pi=5$~GHz to the red sideband of the optical cavity, and a pump laser power of $20$~$\mu$W at the chip. Mechanical modes are identified at frequencies $\omegam/2\pi = (5.1588,5.2146,5.2422,5.2631)$~GHz, with corresponding optomechanical rates $\gom/2\pi = (420,500,527,692)$~kHz. \textbf{b}, Reflected microwave spectrum of the coupled qubit-mechanical system versus qubit bias current tuning.  Microwave signals are measured via the on-chip CPW, which was designed to have direct external coupling rate to the qubit of $\kappaeQ/2\pi= 100$~kHz. Inset shows the qubit's anti-crossing with mechanical mode used for the transducer experiment at $\omegam/2\pi=5.159$~GHz, with a minimum mode splitting of $2g_{pe}/2\pi = 4.5$~MHz}. 
\label{fig:spectrum} 
\end{center}
\end{figure}

Figure~\ref{fig:schematic}b shows the schematic of the transducer region of our device, which consists of a hybridized acoustic cavity formed from a wavelength-scale \piezoacoustic{} resonator connected by a phonon (acoustic) waveguide to an OMC optomechanical cavity. To achieve simultaneously strong electrical and optical coupling to the hybridized cavity modes requires careful design of the individual (i.e., detached) piezoelectric~\cite{Sipahigil2020} and optomechanical~\cite{Chan2012iy} resonators. The \piezoacoustic{} cavity section is designed as a wavelength-scale Lamb wave resonator -- made from a slab of AlN on top of the Si device layer -- that is released from the underlying buried oxide layer and connected laterally to the peripheral substrate via patterned Si tethers that act as acoustic mirrors, minimizing acoustic radiation into the substrate. A pair of aluminum (Al) electrodes in the form of an interdigital transducer (IDT) connect the transmon's capacitive leads to the piezo-acoustic resonator as shown in the schematic of Fig.~\ref{fig:schematic}d. The size parameters of the \piezoacoustic{} resonator are chosen as a compromise between increasing the piezoelectric coupling rate (requiring a large IDT) and minimizing the number of mechanical resonances in the vicinity of the qubit frequency (requiring a small mode volume). The sub-micrometer size scale of the design in Fig.~\ref{fig:schematic}b results in a smaller coupling rate compared to earlier work~\cite{OConnell2010}, but the small mechanical mode density limits the number of parasitic modes which can lead to both qubit decoherence and a reduction in optomechanical coupling. By modifying the OMC cavity, we extend the mechanical energy distribution into the section connecting to the \piezoacoustic{} resonator. This is achieved by deforming the `mirror' cells in the intermediate section to make a transmissive phonon waveguide while maintaining a high reflectivity for the optical field~\cite{Fang:2016ka}. The required spatial extent of this waveguide is set by the design of the optical cavity and the micron-scale wavelength of guided photons. The combined system supports multiple hybridized acoustic modes with different levels of energy concentrations in the AlN and Si sections, one of which is shown in Fig.~\ref{fig:schematic}b. This results in different levels of optomechanical and piezoelectric interaction rates as shown in Fig.~\ref{fig:schematic}c. Although the details of the spectral distribution and interaction rates of the hybridized modes can be effected by geometric disorder introduced during the fabrication process, the strong hybridization via the phonon waveguide gives rise to a robust design, with substantial optomechanical and piezoelectric coupling rates for a range of hybridized modes even in the presence of detuning between the bare \piezoacoustic{} and OMC cavity modes (verified in numerical modelling; See Methods). 
 
The characterization of the qubit and transducer device is performed in a dilution refrigerator, with the sample mounted to the mixing chamber plate of the fridge which is cooled to a base temperature of $\Tf \approx 15$~mK.  Independent sets of optical and microwave spectroscopy measurements are initially performed in order to determine the piezoelectric and optomechanical coupling rates of the hybridized acoustic modes of the transducer.  We perform optical characterization by coupling the light from a tunable external cavity diode laser (ECDL) into a Si waveguide on the chip via a lensed optical fiber (see Fig.~\ref{fig:schematic}e). By measuring the intensity of the reflected light from the on-chip waveguide as a function of laser detuning, the optical modes of the OMC cavity can be determined.  For the device under test in this work, we find an optical resonance at a wavelength of $\lambda=1541.7$~nm with an intrinsic (extrinsic) cavity mode coupling rate of $\kappaoi/2\pi=0.80$~GHz ($\kappaoe/2\pi=0.81$). As shown in Fig.~\ref{fig:spectrum}a, the thermal Brownian motion of the mechanical modes of the OMC cavity (primarily caused by optical absorption heating) can also be observed in the noise power spectral density (NPSD) of the reflected optical signal when measured on a photodiode. The NPSD exhibits a Lorentzian profile for each mechanical mode with a central frequency of $\omegam$ and a linewidth of $\kappam = \kappami + \gammaom$, where $\kappami$ is the intrinsic linewidth of the mechanical mode and $\gammaom \approx 4\ncav\gom^2/\kappao$ is the back-action damping when pumping on the red sideband in the resolved-sideband limit~\cite{Aspelmeyer:2014ce}. Identifying four distinct peaks, we fit the increase in the mechanical linewidth of each mode as a function of optical power to find the optomechanical coupling for each mode (see Methods). The resulting distribution of measured coupling rates can be understood as hybridization of an ideal single OMC cavity mode with $\gom/2\pi \simeq 1.1$~MHz with three additional mechanical modes with zero optomechanical coupling, in good agreement with our design.

We perform microwave spectroscopy of the device using the on-chip coplanar waveguide (CPW) as shown in Figs.~\ref{fig:schematic}d,e. This waveguide has been designed to have direct coupling to the qubit (see Methods), allowing for microwave measurements of the transmon qubit response. Figure~\ref{fig:spectrum}b shows the corresponding measured microwave spectrum, where the qubit is flux tuned using a current biased external coil and the electrically-coupled acoustic modes are identifiable as avoided mode-crossings. The measured acoustic mode frequencies from this experiment are in agreement with the optomechanical spectroscopy measurements, verifying the presence of four hybridized acoustic modes. In particular, the observed acoustic modes at $5.159$~GHz and $5.23$~GHz show hallmarks of strong coupling with the qubit, giving rise to two split branches in the spectrum. Due to the larger value of $\gom$ for the acoustic resonance at $\omegam/2\pi =5.159$~GHz we have chosen this mode (hereafter referred to as the mechanical mode) for the transduction experiments presented below. 

\begin{figure}
\begin{center}
\includegraphics[width=1\columnwidth]{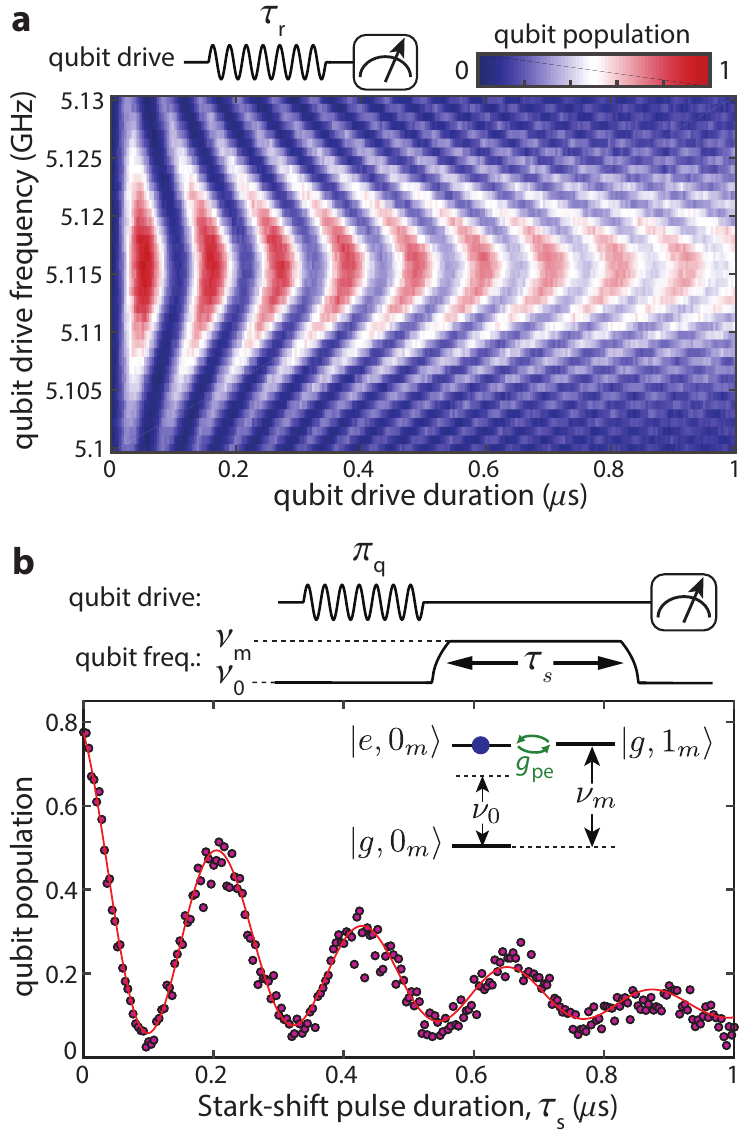}
\caption{\textbf{Qubit-Mechanics Swap. }  \textbf{a}, %
Qubit Rabi oscillations measured using dispersive state readout. \textbf{b}, A $\pi-$pulse ($\tau_r=32$~ns) excites the qubit at frequency $\nu_0 = \omegam/2\pi - 10$~MHz. Tuning the qubit into resonance with the mechanics at frequency $\nu_{\text{m}} = \omegam/2\pi = 5.159$~GHz leads to Vacuum Rabi oscillations between the qubit and the mechanical resonator at frequency $2\gpe/2 \pi =2.24~$MHz. The solid red curve is a phenomenological curve fit consisting of an exponentially decaying sinusoid.  A master equation simulation of the data based upon independently measured decay and dephasing rates for the qubit and mechanical mode yields a qubit-to-mechanics swap time and efficiency of $T_\text{swap} = 104$~ns and $\etaswap = 0.75\pm 0.03$, respectively.}\label{fig:Rabi} 
\end{center}
\end{figure}

Having identified an acoustic resonance strongly coupled to both the qubit and the OMC cavity, we move on to realize a qubit-mechanics swap operation as the initial stage in the microwave-to-optics transduction process. To do this, we first characterize the qubit in the time domain using pulsed excitation through the CPW and dispersive state readout via an integrated microwave read-out (RO) resonator (see Fig.~\ref{fig:schematic}d,e and Methods). Figure~\ref{fig:Rabi}a shows the measured Rabi oscillation of the qubit, which is used to calibrate the duration and magnitude of the microwave drive for realizing $\pi$ and $\pi/2$ pulses. Using calibrated drive pulses, we measure the qubit's free decay profile and Ramsey interference fringes to find the lifetime ($T_{1,\text{q}} = 522\pm 9$~ns) and coherence time ($T_{2,\text{q}}^* = 678\pm 45$~ns) at a bias point ($\omega_q/2\pi = 5.1$~GHz) near the mechanical mode. To perform a coherent qubit-mechanics swap operation, we control the effective interaction time by exciting the qubit, driving it to resonance with the mechanical mode, and tuning it back to the original frequency. Realizing this sequence requires a qubit frequency shift that is several times larger than the interaction rate $\gpe$ and takes place within a transition time that is much shorter than $1/\gpe$. Due to the slow response time of the bias current in the tuning coil, we realize the rapid shift by supplying a sharp pulse ($15$~ns rise/fall time) on the CPW line which is detuned with respect to the qubit by $50$~MHz. This tunes the qubit frequency by $10$~MHz due to the AC-Stark shift, making it resonant with the mechanical mode for the duration of the pulse. Figure~\ref{fig:Rabi}b shows the qubit's excited state population as a function of the interaction time with the mechanical mode, which shows clear evidence of vacuum Rabi oscillations from the piezoelectric coupling. To define a qubit-mechanics swap operation, we find the pulse length of the AC-Stark drive that results in maximal transfer of the qubit excitation to the mechanical mode. From the vacuum Rabi oscillations curve, we find $T_\text{swap} = 104$~ns (in agreement with the value of $\gpe$ from spectroscopy) with a single-phonon initialization probability of $\etaswap = (75 \pm 3)\%$. We have used a pair of swaps with a variable time delay between them to measure the phonon lifetime of the mechanical mode, finding a value of $T_{1,\text{m}} = 357\pm 25$~ns corresponding to an energy decay rate of $\kappamiedecay/2\pi=446$~kHz (this should be compared to the optomechanically measured intrinsic linewidth of $\kappami/2\pi=1$~MHz indicating significant spectral diffusion of the mechanical mode).

\begin{figure*}
\begin{center}
\includegraphics[width=\textwidth]{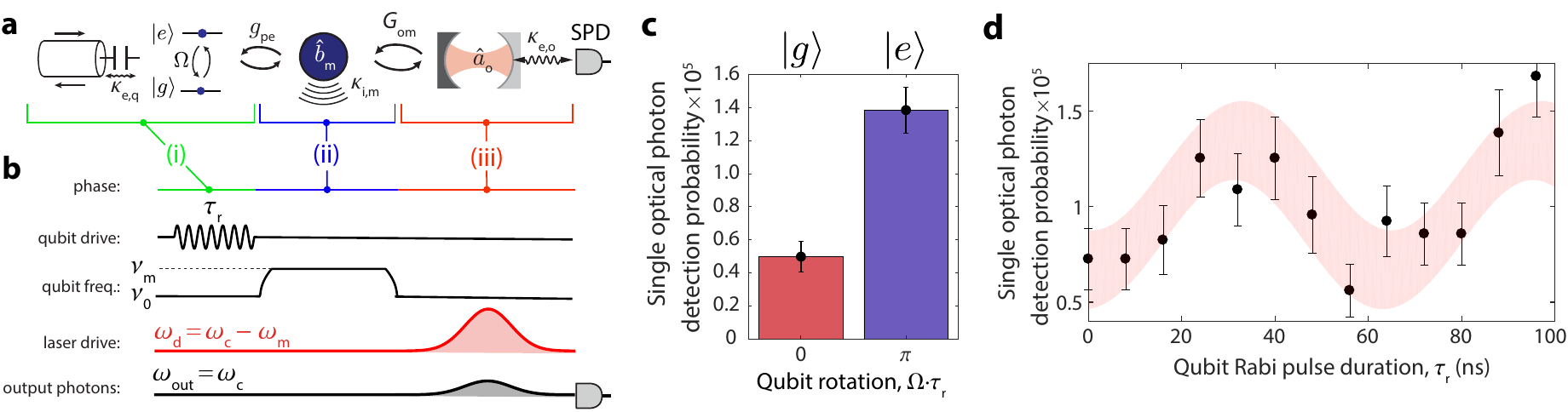}
\caption{\textbf{Detecting optical photons from a superconducting qubit.} \textbf{a-b}, The pulse sequence for quantum transduction. (i) Electrical drive of varying duration excites the qubit. (ii) The resulting qubit state is swapped to the mechanical resonator. (iii) The mechanical state is read out using a red sideband pulse that converts a phonon at frequency $\omegam$ to a an optical photon at frequency $\omegac$. The output photons are detected using a single photon detector (SPD; see Methods). \textbf{c}, $\pi$-pulse probability $\Ppi=(1.38 \pm 0.14) \times 10^{-5}$, no-$\pi$-pulse probability $\Pnopi=(0.50 \pm  0.09) \times 10^{-5}$. \textbf{d}, Qubit Rabi oscillations detected via single optical photon detection. Shaded region shows $90\%$ confidence interval of a sinusoidal fit to the data, with the period of oscillations set to the values from (microwave) dispersive readout of the qubit Rabi curve. Background [$(0.67\pm 0.17) \times {10}^{-5}$] and maximum [$(1.38 \pm 0.17) \times {10}^{-5}$] values from the fit are in agreement with the numbers from (\textbf{c}). Measurements performed using an integration time of $38$~ns with a $10$~ms delay between consecutive pulses (see Methods).}
\label{fig:Transduction} 
\end{center}
\end{figure*}

We now proceed to mechanics-to-optics state transfer as the last stage in the transduction process. We realize this by selectively turning on the `beam-splitter' optomechanical interaction with a red-detuned laser pulse at $\omegad = \omegac - \omegam$. This results in an up-conversion process, in which the phonons in the mechanical mode are mapped to anti-Stokes scattered photons at the center frequency ($\omegac$) of the optical cavity. We use this method to directly measure the phonon occupancy of the mechanical mode by passing the reflected pulsed light from the device through a series of narrow-band tunable filters aligned to the optical cavity (to eliminate the unscattered portion of the pump laser at $\omegad$), and detecting the scattered photons with a single-photon detector~\cite{Cohen:2015hg}. Figure~\ref{fig:Transduction}a,b shows the full transduction sequence which consists of an initial qubit drive by a resonant microwave $\pi$ pulse, followed by a qubit-mechanics swap, and finally the parametric optomechanical conversion of phonons in the mechanical mode to photons. Repeating the experiment with and without qubit drive, we find the scattered photon counts in both cases, which correspond to the excited and ground states of the qubit (see Fig.~\ref{fig:Transduction}c). The data shows a statistically significant separation between the two cases, indicating detection of transduced optical photons from the qubit. We find the overall efficiency and added noise (referred to the qubit) of the transduction scheme to be $\etat = \Ppi - \Pnopi = (0.88\pm 0.16) \times {10}^{-5}$ and $\nadd = \Pnopi/\etat = 0.57 \pm 0.2$, respectively, which are in agreement with the numbers from calibration of the optomechanical readout rate in our system (see Methods). As a further verification, we repeat the experiment while varying the duration of the qubit drive, and observe Rabi oscillations of the qubit population in the detected optical photon counts (see Fig.~\ref{fig:Transduction})d. This provides unambiguous evidence of detecting optical emission from the superconducting qubit by cross-checking the observed period of Rabi oscillations with the results from dispersive qubit readout (Fig.~\ref{fig:Rabi}a). 

Performing optomechanical thermometry without the qubit drive, we have verified a non-zero thermal phonon occupancy as the source of added noise ($\nadd$) in the transduction process (see Methods). The measured residual phonon occupancy can be reduced by using a shorter optical readout pulse, which points to optical absorption heating as the likely origin~\cite{Meenehan2015}. This creates a trade-off in our system, where the noise can be reduced further at the cost of a lower phonon-to-photon conversion efficiency. The measured total read-out efficiency ($\etaRO = \etat/\etaswap = (1.18 \pm 0.26) \times 10^{-5}$) is a product of this intrinsic efficiency ($\etai \approx 10^{-3}$) and the external optical collection efficiency of our measurement set-up ($\etae \approx 10^{-2}$). We note that the measured intrinsic efficiency is significantly lower compared to similar previous optomechanical experiments~\cite{Meenehan2015,Hong:2017id}, due in large part to a faster intrinsic damping rate of the mechanical modes.  The source of this excess mechanical mode damping is currently under investigation.  In addition to efficiency, the repetition rate is another key figure-of-merit. For the current device, the repetition rate of $R=100$~Hz is set by the recovery time of the superconducting circuitry following the application of the readout laser pulse. We have verified optically-induced quasi-particle generation as the main mechanism limiting the repetition rate, and have been able to reduce the recovery time by shortening quasi-particle lifetime using vortex trapping in a cooling magnetic field (see Methods).

Our results mark the first demonstration of optical photon generation from a superconducting qubit. Although we provide direct evidence for transduction of a qubit's excitation, the low photon flux in our experiment does not allow for a direct verification of quantum statistics in the emitted light field. Looking ahead, we identify several avenues for significant improvements in device parameters, which would enable this task and further pave the way for system-level demonstrations of remote entanglement of superconducting qubits. The transduction repetition rate can be improved by more than two orders of magnitude by fabricating qubit electrodes out of materials such as Nb with shorter QP lifetimes~\cite{Johnson1991}. The external read-out efficiency of the experiment can be enhanced by eliminating technical sources of insertion loss in the optical measurement system, leading to nearly two orders of magnitude improvement in the external efficiency. Further, using a material platform with a larger piezoelectric coupling rate would allow for devices with a larger energy participation in the optomechanical cavity, translating to a larger intrinsic read-out efficiency. Finally, a lower added noise is achievable by reducing optical absorption using surface passivisation~\cite{Borselli2006} and by realizing better thermal contact between the mechanical mode and the cold environment~\cite{Ren2020,Qiu2019}. Adoption of these techniques are expected to bring about significant improvements in transduction efficiency ($\etat \sim 0.1$) while maintaining a low added noise value ($\nadd \lesssim 0.1)$. With these improvements, observation of anti-bunching from transduced photons and optically-mediated entanglement generation between remote superconducting qubits should be possible, ultimately leading to new applications for superconducting processors in optical quantum networks.

\begin{acknowledgments}
The authors would like to thank Matthew Shaw, Jash Banker, Hengjiang Ren, Eunjong Kim, and Xueyue Zhang for their various contributions to this work.  This work was supported by the ARO/LPS Cross Quantum Technology Systems program (grant W911NF-18-1-0103), the Institute for Quantum Information and Matter, an NSF Physics Frontiers Center (grant PHY-1125565) with support of the Gordon and Betty Moore Foundation, and the Kavli Nanoscience Institute at Caltech. M.M (A.S.) gratefully acknowledges support from a KNI (IQIM) Postdoctoral Fellowship.  
\end{acknowledgments}

%

\clearpage
\noindent\textbf{\Large Methods}\label{sec:Methods}

\begin{figure*}
\begin{center}
\includegraphics[width=\textwidth]{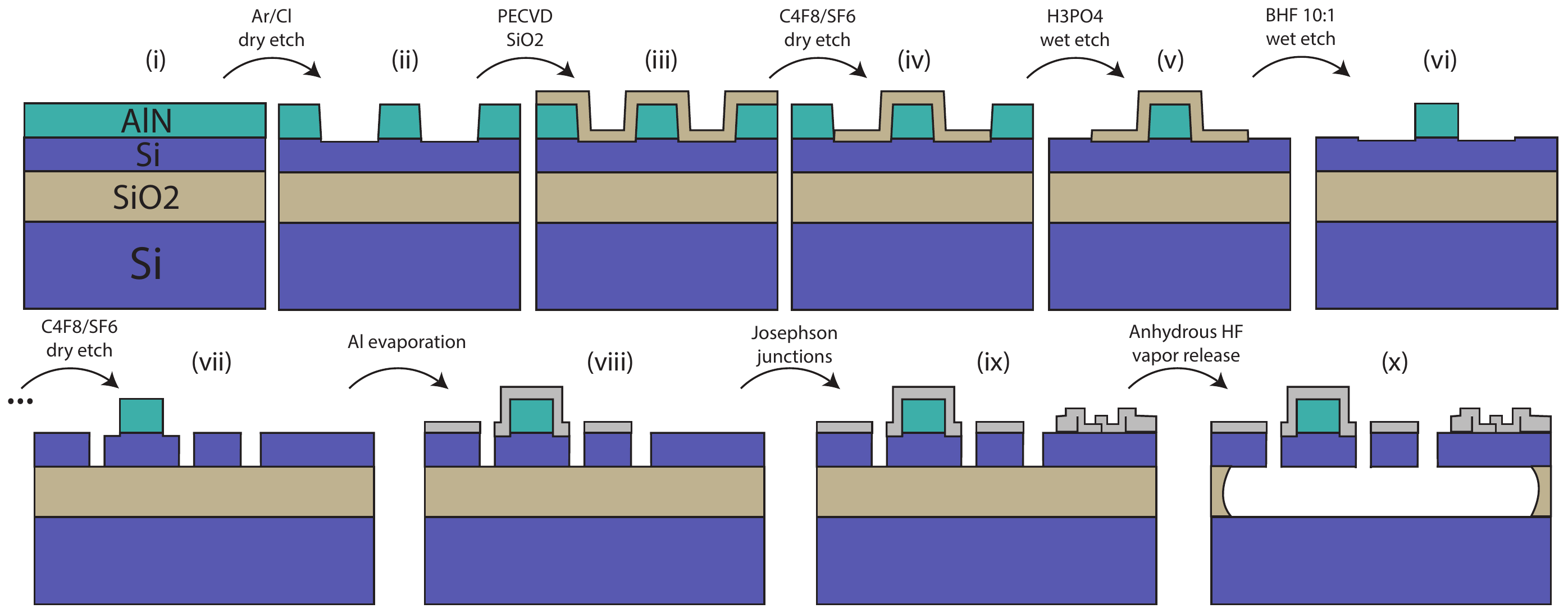}
\caption{\textbf{Device fabrication process.} (i) Starting substrate: AlN-on-SOI. (ii) Defining the boundaries of the piezo-electric resonator with plasma etching. (iii) Deposition of silicon oxide protective mask. (iv) Patterning of the silicon oxide mask with plasma etching. (v) Removing the AlN layer outside the device boundaries with phosphoric acid wet etch. (vi) Removing the silicon oxide mask with a buffered hydrofluoric (HF) acid wet etch. (vii) Patterning of the silicon device layer with plasma etching. (viii) Evaporation of Al for IDT, qubit capacitor, resonator, CPW, and device ground layer. (ix) Angled evaporation of Josephson junctions leads. (x) Vapor-HF etch of buried oxide layer for releasing the device membrane. Images are not to scale. All patterning is performed with e-beam lithography using a ZEP-520A electron beam resist mask.} 
\label{fig:fab}
\end{center}
\end{figure*}

\vspace{1mm}
\noindent\textbf{Fabrication.} The fabrication process for piezo-electric resonator, OMC cavity, and superconducting circuit elements are illustrated in Fig.~\ref{fig:fab}. We start with a $4$-inch silicon-on-insulator (SOI) wafer with the following specifications: Si device layer [float zone grown, $220$~nm thick, $\rho \geq 5$~k$\Omega$-cm]; buried oxide layer [3 $\mu$m thick, silicon dioxide]; Si handle [Czochralski grown, 750 $\mu$m thick, $\rho \geq$ 5~k$\Omega$~cm]. We then perform the following fabrication steps: (i) sputter deposit $300$~nm thick c-axis AlN piezoelectric film (grown by OEM group; stress $T= +55$~MPa; $(002)$ XRD peak of full-width at half-maximum $=1.79^{\circ}$), dice the wafer into $1$~cm $\times~1$~cm chips. The following steps are each carried out using e-beam lithography unless noted otherwise. (ii) Niobium marker deposition (liftoff) and AlN trench etch (Oxford Plasmalab 100 etcher; Ar:Cl$_2$ $=40:80$~sccm; RF power = $120$~W; ICP power $=600$~W; DC Bias $=220$~V ). In this step, we etch only a small area (width $\sim 100$~nm) at the perimeter of the AlN region of the transducer.  This dry etch step ensures that the boundaries of the AlN transducers are precisely defined via dry etching, while still making sure that the silicon device layer is undamaged for the rest of the chip. This is important for achieving  optical, mechanical, and microwave resonances with high quality factors. (iii) Conformal deposition of a hard mask of SiO$_{x}$ via plasma-enhanced chemical vapor deposition.
(iv) patterning of the SiO$_{x}$ mask via dry etching (Oxford PlasmaLab 100 etcher; C$_4$F$_8$:SF$_6$ $=80:40$~sccm; RF power $=14$~W; ICP power $=1300$~W, time $=6$~minutes). (v) Removal of the remaining AlN on the chip with H${}_3$PO${}_4$ (heated to $80~{}^{\circ}$C, $85\%$ by weight). (vi) SiO$_{x}$ mask removal by 10:1 buffered oxide etchant. At the end of step (vi), we have produced a local AlN piezoelectric box (typical dimension $2$~$\mu$m $\times 0.5$~$\mu$m) on the Si device layer while still maintaining a Si surface that is smooth outside the AlN box. The following steps (vii) - (x) follow previously published results for fabricating superconducting qubits on SOI substrates~\cite{Keller2017}. For the devices used in the final experiment in the cryogenic setup, the procedure includes an additional step for realizing end-fire fiber coupling as follows. (xi) We use photo-resist to define a ``trench" region of the chip to be etched for fiber access to the devices' optical waveguides. We use a plasma etch to first remove the buried oxide-layer in the trench region, and subsequently etch the handle silicon in this area to a depth exceeding $100$~$\mu$m. The chips are then cleaned to remove the photoresist and are released in a vapor-HF etch step.

\noindent\textbf{Qubit and microwave readout design.} The superconducting qubits in the experiment are designed to operate in the transmon limit at the resonance frequency of $f_\text{Q} = 5.7$~GHz with capacitive and tunneling junction energies equal to $E_c/\hbar2\pi = 292$~MHz and $E_J/\hbar2\pi = 15.5$~GHz, respectively. This is achieved using a total qubit capacitance of $C_\text{Q} = 66.2$~fF, which includes the IDT contribution along with the junctions' capacitance estimated at $C_J = 4.6$~fF. The junction energy corresponds to a pair of (identical) Josephson junctions in the SQUID geometry with each junction's electrode dimensions designed to be $\AJ = 290$~nm$ \times$ $240$~nm in area. The read-out (RO) resonator is designed as a $\lambda/4$ lumped-element resonator with a resonance frequency of $\fRO = 7.56$~GHz and a (capacitive) coupling rate of $\gRO/ 2\pi = 85$~MHz with the qubit. 

The CPW geometry is designed to achieve an external coupling rate of $\kappaeQ/ 2\pi = 120$~kHz and $\kappaRO/ 2\pi = 5.5$~MHz to the qubit and the readout resonator, respectively. In addition to direct capacitive coupling to the CPW, the qubit can also radiate energy out the CPW port through the non-resonant RO resonator.  The estimated Purcell decay rate for such a process is only $\kappaPQ/2\pi =15$~kHz. 

The qubit and transducers devices were laid out in pairs, with a single CPW designed to provide access to a pair of qubit/RO resonator devices placed in mirror symmetry at the end of the waveguide (see Fig.~\ref{fig:schematic}e). In order to eliminate hybridization between neighboring RO resonators due to near-field coupling (estimated to be $2.5$~MHz from simulations), the pair of RO resonators are intentionally designed to be detuned by $100$~MHz with respect to each other ($7.56$~GHz and $7.66$~GHz). The fabrication disorder in the qubits' resonance frequency ($\gtrsim 50$~MHz from previous experiments and set by variations in junction area) is expected to prohibit hybridization of the neighboring qubits.

\begin{figure*}
\begin{center}
\includegraphics[width=1\textwidth]{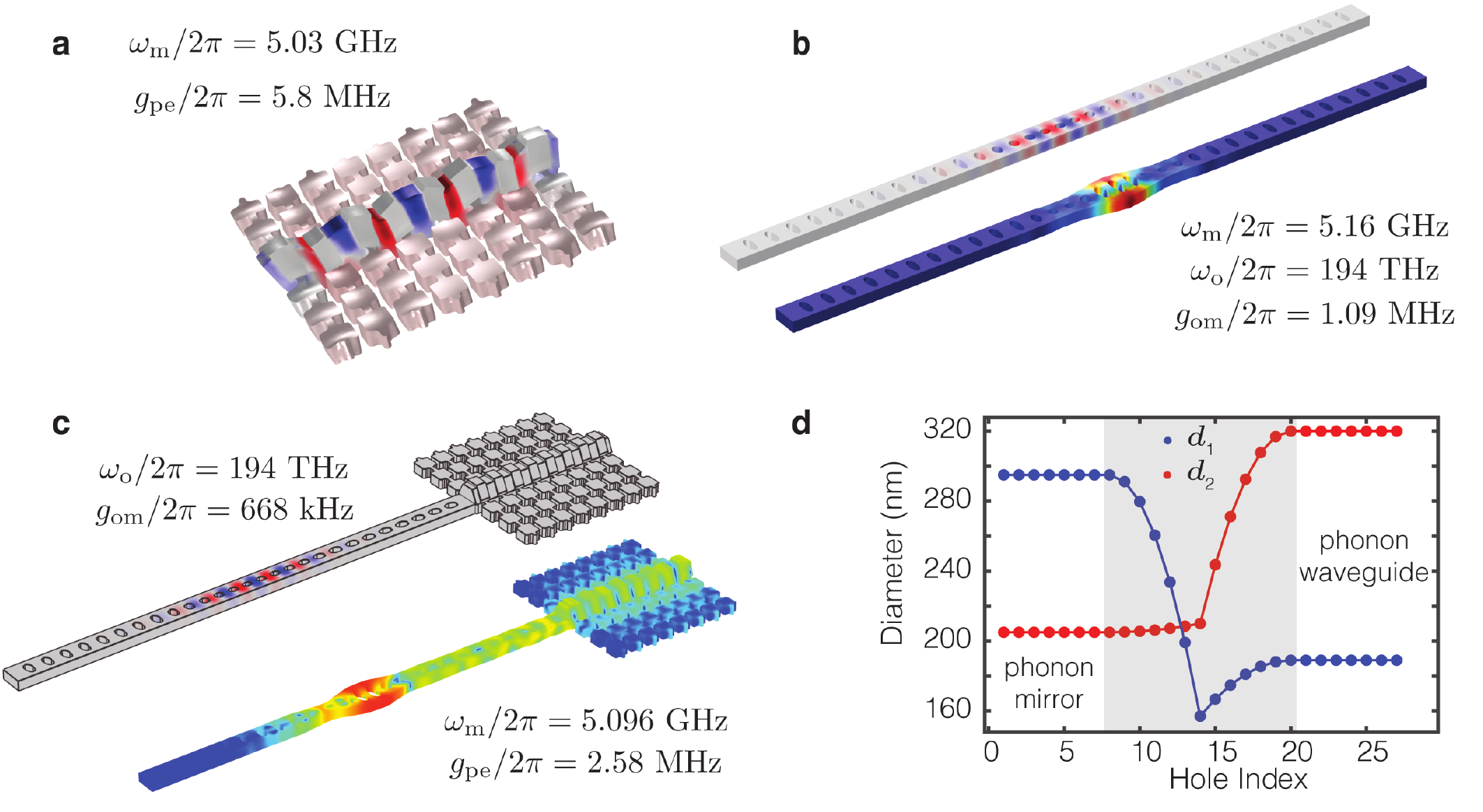}
\caption{\textbf{Optomechanical and piezo-electric design.} \textbf{a}, Simulated mechanical mode shape (deformation) and electric voltage (color) of the \piezoacoustic{} cavity mode of interest with an IDT period of $930$~nm and a beam width diameter of $600$~nm. \textbf{b}, Simulated optical (top) and mechanical (bottom) mode profile of an OMC designed with the mechanical mode near $5$~GHz. \textbf{c}, Simulated optical (top) and mechanical (bottom) mode profile of the full piezo-optomechanical transducer device formed from attaching the \piezoacoustic{} cavity of (\textbf{a}) and the OMC cavity of (\textbf{b}) through a phonon waveguide section in which the mirror holes in the OMC cavity are modified nearest the \piezoacoustic{} cavity. The optomechanical and piezo-electric coupling rates listed are calculated for the hybridized mode with the largest optomechanical coupling. \textbf{d}, Radii of the patterned holes along the nanobeam OMC cavity and phonon waveguide section. $d_1$ ($d_2$) designates the hole diameter normal (parallel) to the nanobeam's long axis. The optical cavity region (shaded $13$ central holes) is located between a phonon/photon mirror (left $7$ holes) and a photon mirror/phonon waveguide section (right $7$ holes).} 
\label{fig:Sims}
\end{center}
\end{figure*}

\noindent\textbf{Optomechanical and piezoelectric simulations.} The optical, mechanical, and piezo-electric properties of the device are simulated using the finite-element method (COMSOL Multiphysics software package)  prior to fabrication. We start with independently designing a pair of nearly-resonant piezo-acoustic and OMC cavities (see Fig.~\ref{fig:Sims}). The piezoelectric coupling rate is extracted by numerical calculation of the overlap integral between the electric field induced by the mechanical mode and the electric field from the qubit applied to the IDT with proper normalization to single-quantum level. To model the piezoelectric response of the AlN thin-film we have used a charge constant that is a factor of 3 smaller than the typically observed bulk value ($d_{33}^\text{bulk} = 4.96$~pm/V), which is found from fitting room-temperature calibration measurements of thin-film AlN-on-SOI piezoelectric transducers fabricated using our process. The optomechanical coupling rate is calculated by numerical evaluation of surface and bulk contributions in a similar fashion to previous work~\cite{Chan2012iy}. In the final design step, we simulate a structure made by attaching the OMC and piezo-acoustic cavity parts and modifying the intermediate section to form a phonon waveguide. The final in-plane dimensions from the simulation are scaled by a factor of $0.96$ prior to device fabrication. This scaling factor is found from calibration measurements for achieving good agreement between the simulated and measured values of optical resonance frequencies of the OMC cavity.

\begin{figure*}
\begin{center}
\includegraphics[width=\textwidth]{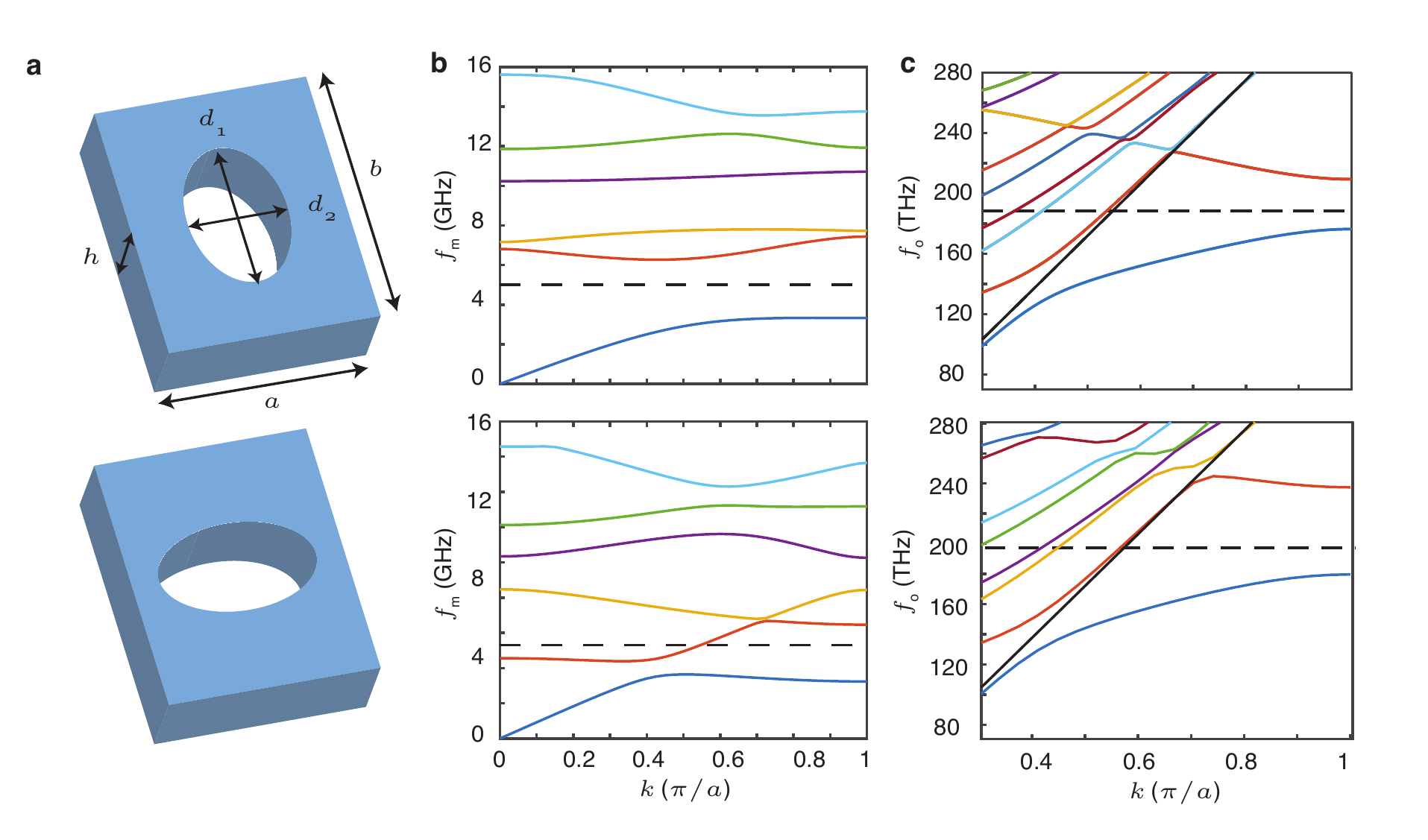}
\caption{\textbf{Design of the phonon waveguide unit cell}. (a) Schematics of the original (top) and modified (bottom) unit cell of the OMC cavity mirror section adjacent to the phonon waveguide. The dimensional parameters are equal to $d_1 = 366$~nm ($d_1 = 295$~nm) and $d_2 = 205$~nm ($d_2 = 320$~nm) for the original (modified) holes. The nanobeam parameters ($h = 220$~nm, $a = 436$~nm, $b = 529$~nm) are identical for both cases. (b) Simulated mechanical band structures. The dashed line marks a nominal mechanical resonance frequency of $f_m = 5.3$~GHz for the decoupled OMC and \piezoacoustic{} cavity modes. The intersection of the dashed line and the energy band in the bottom acoustic bandstructure plot for the modified hole structure allows for guiding of acoustic waves between the \piezoacoustic{} and OMC cavity. (c) Optical band structure for TE-like modes of the OMC cavity, again with the top plot being for the original OMC cavity and the bottom plot for the OMC cavity with modified holes. The solid black line marks the light line and the dashed line refers to a nominal optical resonance frequency of the fundamental mode of the OMC cavity at a frequency $f_o = 193$~THz.  Unlike the acoustic mode case, the hole modifications in the OMC cavity actually increase the optical bandgap, further suppressing optical radiation into the phonon waveguide.} 
\label{fig:PhononWG}
\end{center}
\end{figure*}

\noindent\textbf{Phonon waveguide design.} Mechanical hybridization between the AlN-on-SOI \piezoacoustic{} resonator and the Si OMC cavity is achieved by modifying the OMC structure in order make a mechanically-transparent section between the two components. Figure~\ref{fig:PhononWG} shows a unit cell of the nanobeam in this intermediate section in the original OMC design and after modifying it. In order to make this section transparent to mechanical waves, we have simulated the optical and mechanical band structure of the unit cell while sweeping the ellipticity of the central `hole' in the nanobeam. Heuristically, this modification results in a unit cell with a similar dielectric to vacuum ratio (due to the nearly constant area of the holes), which maintains the qualitative form of the optical band structure, whereas the acoustic band structure is significantly modified due to the hard boundary between Si and vacuum for acoustic waves (acoustic waves can't travel in vacuum as electromagnetic fields do).  

\begin{figure*}
\begin{center}
\includegraphics[width=\textwidth]{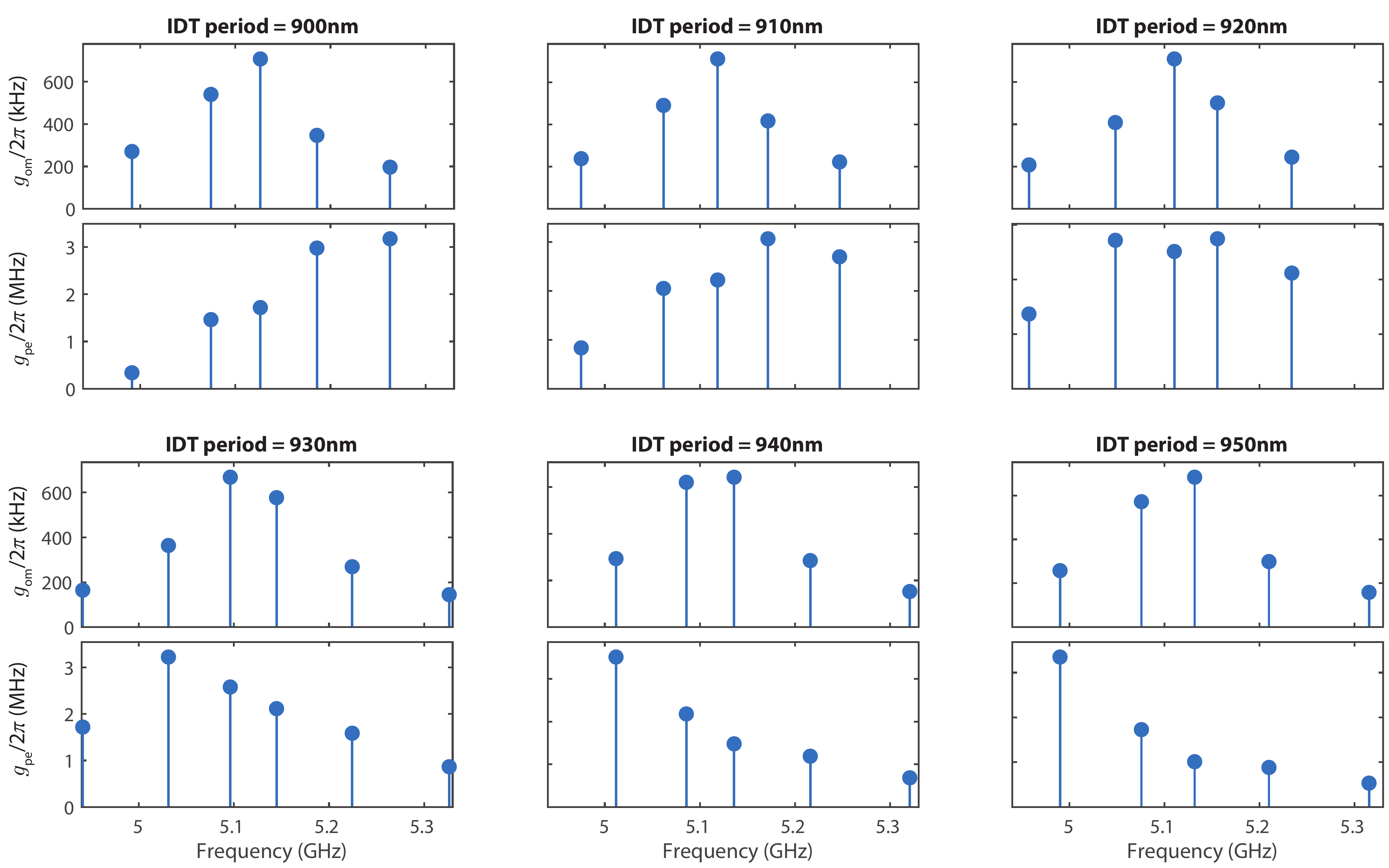}
\caption{\textbf{Mechanical mode hybridization}. Optomechanical and piezoelectric coupling rates from COMSOL Multiphysics simulations of the integrated piezo-optomechanical transducer consisting of a \piezoacoustic{} cavity connected by a phonon waveguide to the OMC cavity (see Fig.~\ref{fig:schematic}b). Each resonant acoustic mode of the simulated structure is plotted as a `pin' in each plot, with the horizontal axis corresponding to the resonant mode frequency, $\omegam/2\pi$. The period of the IDT has been varied along with the total length of the \piezoacoustic{} cavity (i.e., fixed total number of periods and exterior boundary region to AlN box), leading to an approximate linear change in the frequency of the mechanical modes with IDT period. The OMC cavity and phonon waveguide parameters are identical for all of the plots.} 
\label{fig:MechHyb}
\end{center}
\end{figure*}

\noindent\textbf{Disorder and acoustic mode hybridization.} The phonon waveguide design presented in the previous section allows for achieving hybridization between the OMC and \piezoacoustic{} cavity modes when the two modes are in resonance. In practice, however, the exact resonance frequency of each component cavity is dependent on variations in the geometry that inevitably happen during the fabrication process. This variation can be reduced by careful calibration of the fabrication process to values $\lesssim 50$~MHz. Alternatively the devices can be trimmed post fabrication to achieve the resonance conditions. In our experiment, we eliminate the need for such measures by realizing a large mechanical inter-mode coupling between the OMC and the piezoelectric resonator via the phonon waveguide, which makes the hybridization a weak function of the detuning between the two. Figure~\ref{fig:MechHyb} shows the simulated optomechanical and piezoelectric coupling rates of the hybridized modes for different values of the IDT period in the \piezoacoustic{} cavity. Assuming a linear dispersion for the mechanical modes confined in the \piezoacoustic{} cavity, the $50$~nm change of the period in Fig.~\ref{fig:MechHyb} corresponds to approximately a $280$~MHz change in the detuning between the OMC and \piezoacoustic{} cavity modes. As evident from simulations, while the details of the spectral composition and coupling rates is subject to change, the mechanical mode hybridization is robust to non-zero detunings resulting in qualitatively similar mode spacing and coupling rates for various IDT periods. For the fabricated devices, we have swept the IDT period by $20$~nm ($\sim 110$~MHz frequency shift) around the central periodicity of $885$~nm.

\begin{figure*}
\begin{center}
\includegraphics[width=0.85\textwidth]{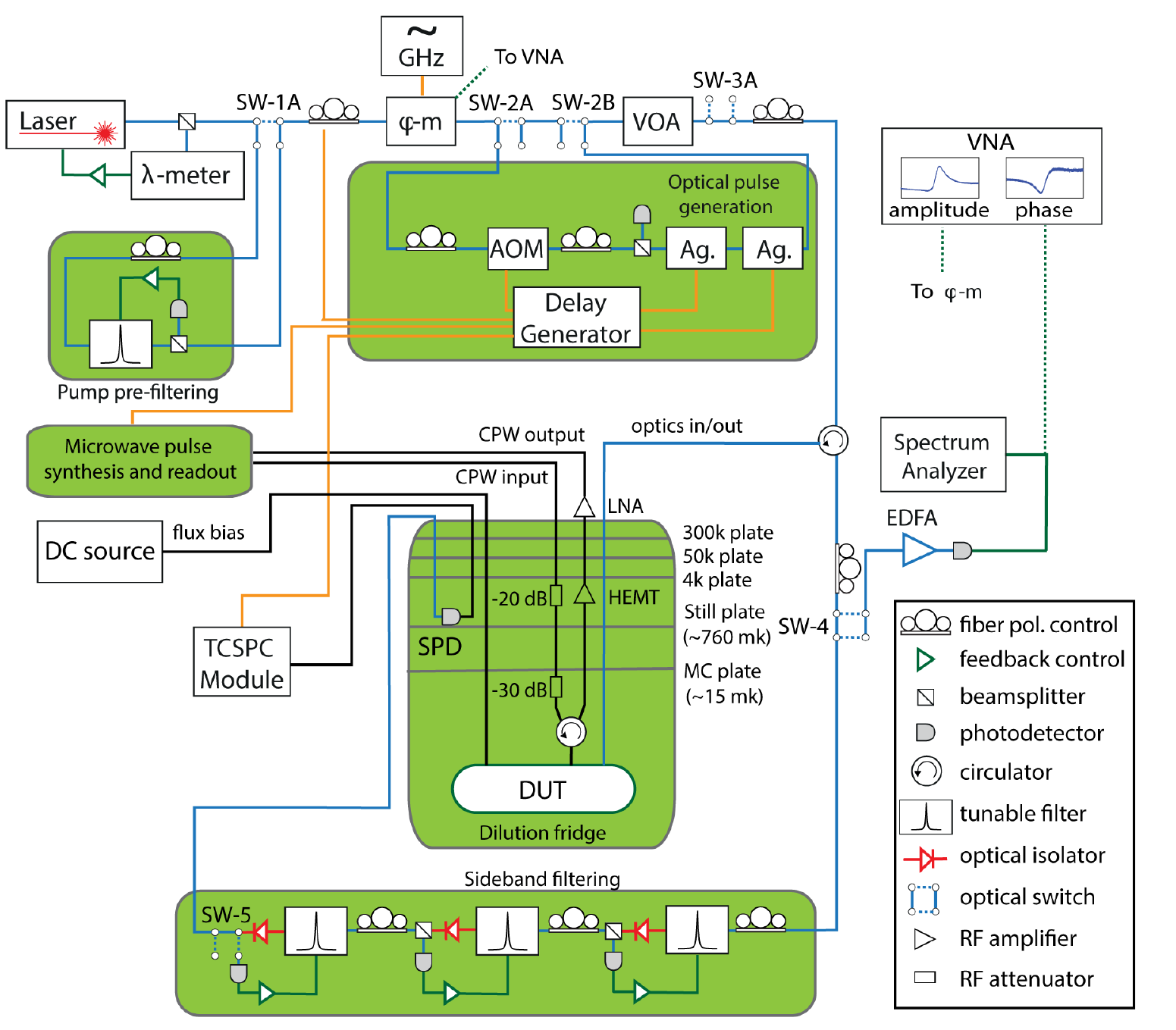}
\caption{\textbf{Transduction measurement setup}. Simplified diagram of the experimental setup used for low-temperature piezoelectric and optomechanical device characterization. The laser emission is passed through a $50$~MHz-bandwidth filter to suppress broadband spontaneous emission noise. The optical readout pulses are generated using high-extinction modulation components (AOM, Ag.). Microwave pulses are synthesized using arbitrary waveform generators at intermediate frequency band (IF, $<500$~MHz) and up-converted using mixers and microwave sources to the GHz frequency band. Readout of microwave signals is performed using an analog-to-digital converter after down-conversion to IF band. The optical modulation and microwave synthesis/readout components are triggered by a digital delay generator. Upon reflection from the device under test (DUT), a fiber optical circulator routes the outgoing light to either (1) an EDFA and spectrum analyzer, or (2) a sideband-filtering bank consisting of three cascaded fiber Fabry-Perot filters (Micron Optics FFP-TF2) and the SPD operated at $\sim 760$~mK. Electrical (optical) signal/control line shown in black (blue). $\lambda$-meter: wavemeter, $\phi$-m: electro-optic phase modulator, AOM: acousto-optic modulator, Ag.: Agiltron 1x1 MEMS switch, SW: optical $2\times2$ switch, VOA: variable optical attenuator, EDFA: erbium-doped fiber amplifier, VNA: vector network analyzer, SPD: single photon detector, TCSPC: time-correlated single photon counting module (PicoQuant PicoHarp 300). } 
\label{fig:MeasSetup}
\end{center}
\end{figure*}

\noindent\textbf{Measurement setup.} Figure~\ref{fig:MeasSetup} shows a schematic of the measurement setup. The arrangement of major optical components is adopted and modified from previous work on phonon counting \cite{Cohen:2015hg,maccabe2019phononic}. A digital delay generator is used to synchronize the generation of the microwave drive, the optical readout pulses generated by optical modulation, and the placement of the detection window in time. Optomechanical readout and characterization is performed with an ECDL source that is frequency stabilized using a wavemeter in a computer feedback loop. The laser light is pre-filtered to avoid laser phase-noise at the mechanical resonance frequency. A phase modulator is used to create optical sidebands for the purpose of coherent driving of the mechanical modes (during characterization) or for creating an optical reference for locking the tunable filter array in the detection path (as part of the transduction sequence). Readout optical pulses are shaped using an acousto-optic modulator (rise and fall time of $20$~ns) and a pair of mechanical switches (rise/fall time of $100$~ns/$30$~$\mu$s) cascaded together to achieve high-contrast pulses (extinction $>86$~dB). The reflected optical signal from the device is routed via an optical circulator to a mechanical switch, where it is directed towards two separates paths for performing continuous-wave and pulsed measurements. 

For characterization of the acoustic modes of the OMC cavity, the optical reflected signal is routed to the path with an erbium-doped fiber amplifier (EDFA) and a high-speed photodetector (PD). The photo-current from the detector is registered with a analog-to-digital converter to measure the optical resonance while sweeping the laser frequency. Measurement of the thermal Brownian motion of the mechanical modes is performed using a spectrum analyzer (SA) for registering the photo-current when the laser is locked at $\Delta \approx \omegam$. The coherent mechanical response of the device is measured with two-tone spectroscopy, where the pump laser light is modulated with a phase-modulator driven by the output of a vector network analyzer (VNA), and the RF component of the photo-current is measured at the input port of the VNA. For the transduction experiment, the second detection path is used for photon counting. Here, the light is passed through three cascaded high-finesse tunable fiber Fabry-Perot filters (Micron Optics FFP-TF2) placed inside a thermally insulating housing, and is then routed to a single-photon detector (SPD) mounted on the still plate inside the dilution refrigerator. The SPD used in the experiment is a WSi-based superconducting nanowire detector. The RF output of the SPD is amplified with a cryogenic amplifier at the $50$~K stage of the dilution fridge and a room-temperature amplifier prior to detection with a triggered time-correlated single photon counting (TCSPC) detection module.

The fabricated device is cooled to $\Tf \approx 15$~mK in a dilution refrigerator, and is shielded from the magnetic environment using a mu-metal shield inserted into the external vacuum can of the cryostat and a cryoperm shield located at the mixing chamber plate.  Optical alignment and coupling of laser light to the tapered Si optical waveguide of a given transducer device is performed using a stack of cryogenic piezo steppers. Microwave signals are routed to the device via a coaxial cable, with thermal grounding to the 4K and mixing chamber plates of the fridge, and then onto the CPW of a printed circuit board which is wire bonded to an on-chip CPW. Reflected microwave signals from the device are collected using an RF circulator and amplified through an amplifier chain consisting of a high electron mobility transistor amplifier (HEMT) at the 4K stage of the fridge and a low noise amplifier (LNA) at room-temperature. Qubit frequency tuning is achieved by flux tuning of the SQUID loop using an external hand-wound coil made from Nb-Ti superconducting wire which is placed several millimeters above the chip. The flux bias current is applied to the coil using a low-noise dc source.

\noindent\textbf{Optomechanical scattering rate and mechanical mode occupancy calibration.} We measure the spectral response of the mechanical modes by performing a two-tone spectroscopy technique. In this approach, the laser is locked to the red sideband of the optical cavity ($\Delta = \omegam$), and phase modulation is used to generate a pair of optical sidebands, where one of the sidebands is swept across the optical cavity. For the case where the pump-sideband frequency separation coincides with the mechanical resonance frequency, the optical susceptibility of the cavity is strongly suppressed due to cancellation of coherent scattering components from the mechanical and optical resonances, in a similar fashion to electromagnetically-induced transparency\cite{Meenehan2015}. We calibrate the optomechanical coupling rate of the mechanical resonance at $\omegam/2\pi = 5.159$~GHz by fitting the linewidth from the two-tone spectroscopy as a function of the pump photon number in the optical cavity, finding $\gom/ 2\pi = 420$~kHz. The optomechanical scattering rate for other mechanical resonances are found by integrating the area under the thermal Brownian spectrum at each resonance frequency (measured using a spectrum analyzer) and using the optomechanical coupling rate of the mode at $\omegam/2\pi = 5.159$~GHz as a reference.

The optomechanical coupling rate is independently verified by measuring the photon scattering rates with the pump locked to the red and blue sideband of the optical cavity ($\Delta = \pm \omegam$). In this setting, the photon detection rate can be written as,

\begin{align}
\Gamma_{\text{R}/\text{B}}(t) = 
& \Gamma_\text{dark} + \eta_\kappa \etasys  \frac{4\gom^2\ncav(t)}{\kappao} \left(\nm + \frac{1}{2} (1\mp 1) \right), 
\end{align}

\noindent where $\Gamma_\text{dark} = 10$~cps is the rate of dark counts, $n_\text{m}$ is the occupancy of the mechanical mode, and $\eta_\kappa = \kappa_\text{e,o}/(\kappa_\text{i,o}+\kappa_\text{e,o})\simeq 0.5$ is the cavity-to-waveguide coupling efficiency. The external optical efficiency of the system is independently measured to be $\etasys = 1.5\%$, which includes the efficiency of coupling from the lensed fiber to the device ($\etacoupler = 65\%$, one-way), transmission efficiency through the measurement system including the optical filter bank ($\etatran = 3\%$), and the quantum efficiency of the SPD ($\etaSPD=85\%$). 

So long as the optomechanical back-action rate is negligible compared to the intrinsic mechanical mode damping rate (i.e., $\nm$ is not influenced by the back-action), the difference in red- and blue-sideband scattering rates provides a calibration-free means of determining the per-phonon count rate by using the vacuum contribution (i.e., spontaneous Stokes scattering) as a reference.  From the independent measurements of $\gom$ and $\kappao$ we estimate that for the optical drive power used in these experiments ($2$~$\mu$W at the chip; $\ncav=44$) that the optomechanical back-action is $\gammaom/2\pi = 19$~kHz, far less than the measured $\kappamiedecay/2\pi = 446$~kHz.   We can use this to find an independent estimate of the optomechanical readout efficiency of the transducer.  The optomechanical read-out efficiency as a function of read-out time, $\tauRO$, is given by,

\begin{equation}
\etaRO(\tauRO) = \left(\frac{\gammaom}{\gammaom + \kappamiedecay}\right)\left(1 - \exp{[-(\gammaom+\kappamiedecay)\tauRO]}\right),
\label{eq:etaRO}
\end{equation}

\noindent which in the long-time limit is just the fraction of `good' damping, $\gammaom/(\gammaom+\kappamiedecay)$. In the small-time limit ($\tauRO \ll 1/(\gammaom+\kappamiedecay)$), appropriate to the measurements reported here, the optomechanical readout efficiency can be simply related to the difference in the blue and red sideband scattering rates, $\etaRO(\tauRO) \approx p_d \equiv \int_0^{\tauRO}(\Gamma_\text{B}(t) - \Gamma_\text{R}(t))\text{d}t$, where the integration accounts for the fact that the scattering rates may be time-dependent due to the turn-on of the optical pulse. Similarly, we find the time-averaged (over $\tauRO$) residual thermal noise of the heated mechanical mode occupancy by integrating the red-sideband rate normalized to the per-phonon rate: $\langle \nm \rangle_{\tauRO} = \int_0^{\tauRO}\Gamma_\text{R}\text{d}t/\int_0^{\tauRO}(\Gamma_\text{B} - \Gamma_\text{R})\text{d}t$.

\begin{figure*}
\begin{center}
\includegraphics[width=0.55\textwidth]{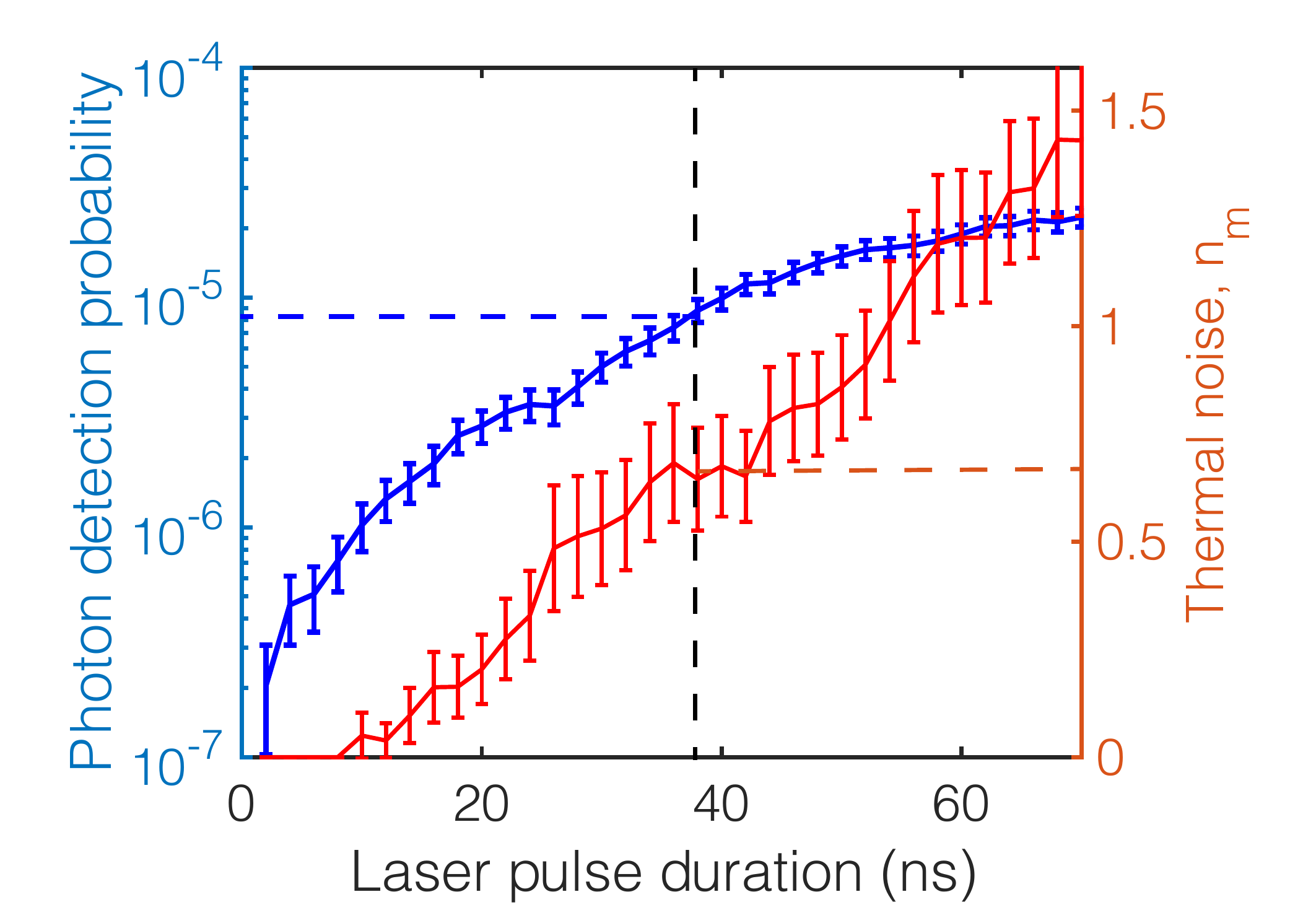}
\caption{\textbf{Laser-induced heating}. Plot of the measured thermal noise in the mechanical mode and optomechanical read-out efficiency as a function of integration time ($\tauRO$).  Here the noise and efficiency are calibrated from measurements of the optomechanical sideband asymmetry.  As for the transduction measurements, the optical pump power is set to $P_{\text{in}} = 2$~$\mu$W at the optical input fiber to the dilution refrigerator, which corresponds to an estimated intra-cavity photon number of $n_c=44$ and an optomechanical readout rate of $\gammaom = 4\gom^2/\kappa_\text{o}=2\pi\times [19~$kHz]. The measurement sequence contains no microwave drive, with the qubit detuned from the mechanical mode, and contains a $250~ \mu$s delay between consecutive optical readout pulses to ensure relaxation of the mechanical mode occupancy to its base-temperature value ($\ll 1$) with the laser off. The error bars mark the standard deviation at each point, which is calculated using the raw counts assuming Poissonian shot noise.  For comparison to the transduction measurements presented in Fig.~\ref{fig:Transduction}, the operational readout window of $\tauRO = 38$~ns yields a photon detection efficiency of $p_d= 8.8 \pm 1.3 \times 10^{-6}$ and a thermal noise of $n_\text{m} = 0.64 \pm 0.15$ phonons.} 
\label{fig:heating}
\end{center} 
\end{figure*}

Figure~\ref{fig:heating} shows the calculated noise and efficiency from the optomechanical scattering measurements as a function of the pulse integration time ($\tauRO$) with the qubit decoupled from the transducer. As is evident, the time-averaged noise phonons in the mechanical resonator starts with a small value and grows with the optical pulse measurement time, in agreement with previous observations of optical absorption heating in silicon OMC devices.  The readout efficiency also increases with integration time, leading to a trade-off between noise and efficiency. For an optical read-out integration time of $\tauRO = 38$~ns, equivalent to that used in the qubit transduction data of Fig.~\ref{fig:Transduction}, the extracted average mechanical mode occupancy and optomechanical readout efficiency are $\langle \nm \rangle_{\tauRO} = 0.64 \pm 0.15$ and $\etaRO(\tauRO) = (0.88 \pm 0.13) \times 10^{-5}$, respectively.  These 'vacuum-calibrated' values are consistent with the corresponding `qubit-calibrated' values of $\etaRO^{q} = (1.18 \pm 0.26) \times 10^{-5}$ and $\langle \nm \rangle^{q} = \nadd\etaswap = 0.43 \pm 0.17$.  

\begin{figure*}
\begin{center}
\includegraphics[width=0.6\textwidth]{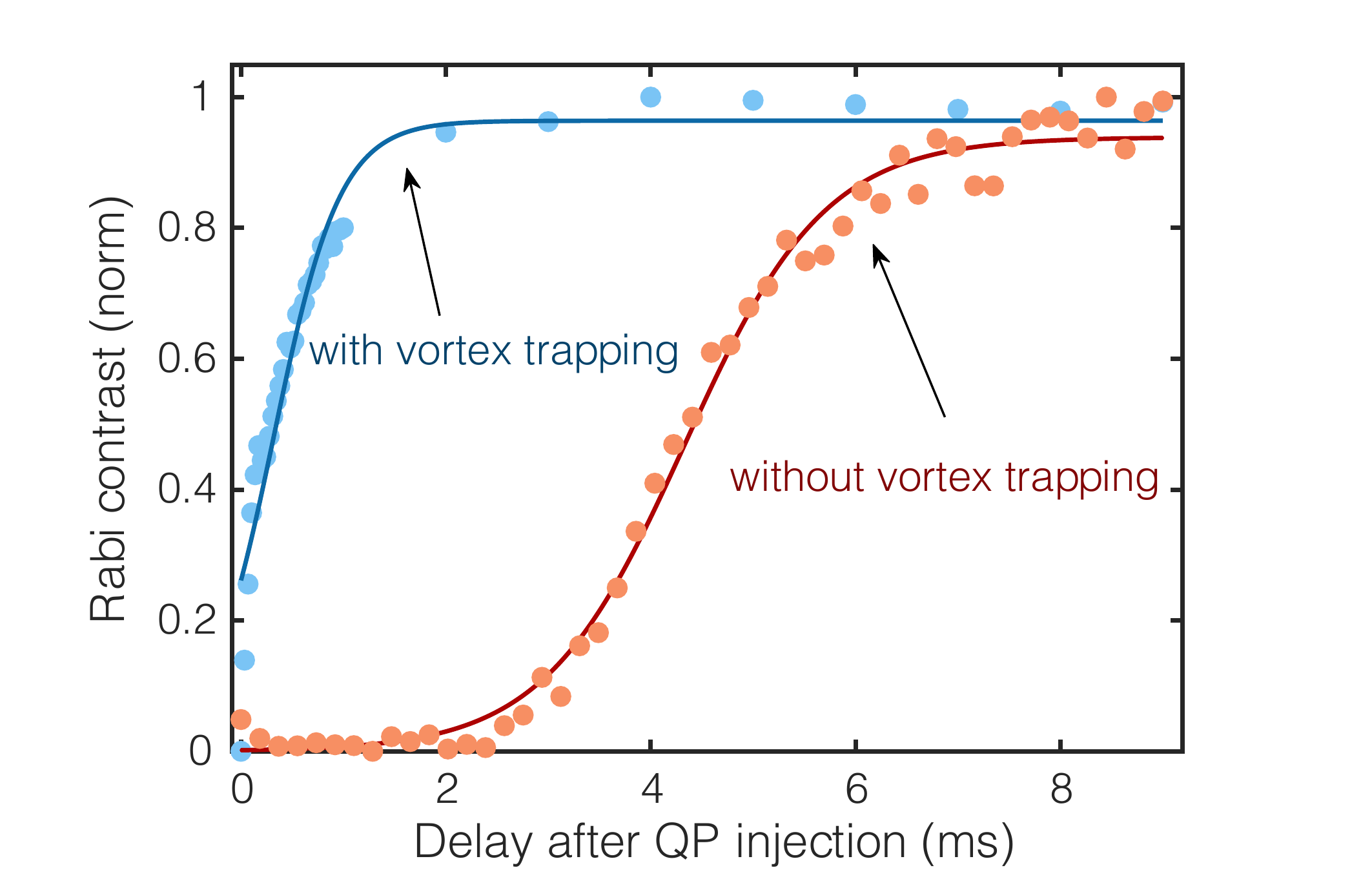}
\caption{\textbf{Light-induced quasi-particle generation}. Plot of the qubit Rabi contrast measured when the transmon qubit is driven with a variable-length microwave drive after illumination by an optical pulse (Rabi oscillation period $150$~ns). $X$-axis marks the separation between incident optical pulse and the beginning of the Rabi measurement. The optical pulse duration is $\tau_{\text{pulse}}=100$~ns and the pulse repetition rate is $R=10$~Hz. The peak optical power of the pulses corresponds to $P_{\text{in}} = 40$~$\mu$W at the input fiber to the dilution refrigerator. Vortex trapping is done with a cooling magnetic field estimated to $15$~Gauss at the chip.} 
\label{fig:QPgen}
\end{center}
\end{figure*}

\noindent\textbf{Quasi-particle trapping.} Absorption of pump laser light in our experiment leads to generation of excited electrons (i.e., quasi-particles) from broken cooper pairs due to the large energy of the infrared photons with respect to the energy gap of superconducting aluminum $\hbar \omegac \gg 2\Delta_\text{gap}$. The generated non-equilibrium quasi-particle (QP) population in the transmon's capacitive leads gives rise to a dissipative component in the admittance of Josephson junctions, which reduces the lifetime of the qubit, and in turn compromises the ability to perform coherent gate operations in the microwave domain. This excess quasi-particle population eventually decays to the steady-state value via electron-phonon-mediated pair recombination with a characteristic lifetime set by the material properties of the superconducting film and the substrate~\cite{Wang2014}. Figure~\ref{fig:QPgen} shows the QP relaxation measured by sending optical pulses into the transducer device, followed by a delayed measurement of the qubit Rabi curve contrast performed via the microwave readout resonator. As is evident, for the pulse duration and optical power used in the transduction measurements, the transmon qubit requires approximately $8$~ms delay between pulses to allow for full recovery (hence our limited repetition rate in the transduction measurements of $R=100$~Hz).

To verify QP generation as the source of light-induced decoherence, we have performed electrical QP relaxation measurements, where we inject QPs into the qubit by creating a large off-resonant AC voltage across the qubit junctions~\cite{Wang2014} via a pulsed drive applied on the CPW. Using the measured decay rate of the Rabi oscillations of the qubit as an indicator of the qubit's coherence time after the QP injection with a variable delay, we extract a QP lifetime of $1.5$~ms for our device. The measured value is in qualitative agreement with previously reported values for thin-film aluminum and is qualitatively consistent with results of the optics measurement. In the next step we try to reduce QP relaxation lifetime via vortex trapping~\cite{Wang2014}. To do this the device is warmed to above the superconducting transition temperature, a magnetic field (normal to substrate) is applied via the flux-tuning coil and the device is subsequently cooled down to base temperature. This results in creation of vortices in electrodes with the number of vortices approximately equal to the magnetic flux threading a square of electrode width divided by flux quantum. At the base temperature, we turn off the magnetic field slowly, which is expected to pin the vortexes to defects in the thin film aluminum. We have then repeated the electrical measurement of QP relaxation, finding a lifetime of $320$~$\mu$s for a cooling magnetic field of $15$~Gauss (corresponding to $\sim 75$ vortices per square in the transmon electrodes). As a final verification, we repeat the light-induced QP generation experiment in this setting, finding a recovery time of approximately $2$~ms for the transmon qubit. The electrical and optical measurements consistently point to a factor of $\sim 5$ enhancement in the measurement repetition rate via vortex trapping of QPs.

\end{document}